# WILLINGNESS TO PAY FOR BASIC RESEARCH:
# A CONTINGENT VALUATION EXPERIMENT
# ON THE LARGE HADRON COLLIDER

Gelsomina Catalano[1], Massimo Florio[1], Francesco Giffoni[2]

[1] Department of Economics, Management and Quantitative Methods, University of Milan

[2] CSIL Centre for Industrial Studies

This version: 10.03.2016

**Abstract**

An increasing number of countries and institutions are investing in large-scale research infrastructures (RIs) and in basic research. Scientific discoveries, which are expected thanks to RIs, may have a non-use value, in analogy with environmental and cultural public goods. This paper provides, for the first time, an empirical estimation of the willingness to pay (WTP) for discoveries in basic research by the general public. We focus on the Large Hadron Collider (LHC), the largest particle accelerator worldwide, where in 2012 the Higgs boson was discovered. Nobody knows the practical value of such discovery, beyond knowledge per se. The findings of our study are based on a dichotomous choice contingent valuation (CV) survey carried out in line with the NOAA guidelines. The survey involved 1,022 undergraduate students enrolled in more than 30 different degrees (including the humanities) at five universities located in four countries (Italy, France, Spain, UK). We ask two main research questions: Which are the determinants of the WTP for the LHC discoveries? What is the average contribution that the respondents would be willing to pay? Our results confirm that income, interest and attitudes towards basic research are positively associated with the WTP, while other potential explanatory variables play a limited role. The estimated mean of WTP for basic research in particle physics is EUR 7.7 per person una-tantum. Although this is a small amount compared to other CV studies for environmental and cultural goods, it points to positive social attitudes for basic science as a public good.

**Acknowledgments**: This paper has been produced in the frame of the research project '*Cost/Benefit Analysis in the Research, Development and Innovation Sector*' sponsored by the EIB University Research Sponsorship programme (EIBURS), whose financial support is gratefully acknowledged. Further details on this research project can be found at: http://www.eiburs.unimi.it/

The authors are grateful to Ana Almuedo-Castillo (Exeter University), David Attié (Paris), Gianluigi Magnetta (University of Milan), Paulino Montes (University of A Coruña), Chiara Pancotti (CSIL) for collaboration in interviews. They are also grateful to Emanuela Sirtori and Silvia Vignetti for valuable comments.

# 1. Introduction

In the last decades, an increasing number of governments and institutions have supported basic research defined by OECD (2002) as 'experimental or theoretical work undertaken primarily to acquire new knowledge of the underlying foundation of phenomena and observable facts, without any particular application or use in view'. In the near future, several and ambitious projects are at stake. The 'Europe 2020' strategy includes the Innovation Union flagship initiative, aimed at transforming the EU area into a world-class science performer by completing or launching the construction of priority European RIs (European Commission, 2010). Within this framework, the self-regulatory body European Strategy Forum on Research Infrastructures (ESFRI) aims at developing a joint vision and a common strategy including updated roadmaps, reports and criteria as tools for planning and implementing new pan-European RIs.[1] Along the same line and after the demise in 1993 of the Superconducting Super Collider (SCC) in the US[2], the CERN in Geneva has achieved the leadership in particle physics thanks to the Large Hadron Collider (LHC), where the Higgs Boson was discovered in 2012. Recently, the CERN has also launched a study for the Future Circular Collider (FCC), laying the foundations for the post-LHC era (Reich, 2013; Banks, 2014).[3] Other countries, including China, are planning large-scale scientific ventures for the next decades.[4]

These projects are very costly. In 1993, the US Congress stopped the SSC project because of an increase in the estimated costs from 4.4 billion USD to 12 billion USD.[5] The present value to 2025 of the LHC is around 13.5 million EUR (Florio *et al.*, 2016a). Is it worth for the taxpayers to fund such projects? The question is particularly intriguing when basic research is considered, given that its definition acknowledges that it has yet 'no use'.

In the traditional perspective of welfare economics, the value of a good arises from its use, or utility. Nevertheless, since the Sixties environmental economists have been arguing that there may be a value arising from its non-use, including the pure existence of the good itself (Weisbrod, 1964; Krutilla, 1967). Bateman *et al.* (2002) classify the non-use value into three main categories: the bequest value, the option value and the existence value; in some cases the notion of quasi option value is added (Boardman *et al.*, 2001). The option value arises when it is possible to predict some use of the good in the future but there is not yet a use at the present time. If the future use of the good is not known yet or it is today unpredictable, and there is irreversibility, then the notion of quasi-option value has been proposed.[6] Differently from bequest (or altruism), option and quasi option values, the existence value originates from the utility that arises from the mere perception of the existence of the good, even in the absence of any expected or unpredictable use (Walsh *et al.*, 1984; Brun, 2002). Several studies worldwide provide empirical estimations of the existence value of environmental goods (among others, Heafele *et al.*, 1991; Chopra, 1993; Echeverria *et al.*, 1995; Loomis *et al.*, 1996; Costanza *et al.*, 1997; White and Lovett, 1999; Amirnejad *et al.*, 2006). During the last twenty years, the concept of existence value has been transferred to cultural economics (Hansen, 1997; Frey, 2000; Alberini *et al.*, 2003; Alberini and Longo, 2006; Packer, 2008; Marsh *et al.*, 2010; Fujiwara, 2013).[7]

Florio and Sirtori (2015) and Florio *et al.* (2016b) suggest that the notion of existence value can be extended to the RIs, and, in general, to science as well. Specifically, the authors argue that there may be a social preference for new knowledge *per se*, and as a consequence, the existence value of RIs should arise from the pleasure (or utility) of knowing that something may be discovered even if there is no predictable use of it. If so, scientific discoveries can be easily associated with environmental goods: the only difference is that natural environments are something that are known to exist and that may be endangered, while a discovery reveals something that already exists, but was previously or still unknown. The analogy holds for cultural goods as well, if the preservation of cultural heritage is valued in itself and not because of its market value due, for example, to tourism.

---

[1] See for details http://ec.europa.eu/research/esfri
[2] See http://www.scientificamerican.com/article/the-supercollider-that-never-was/
[3] The FCC will be a proton-colliding machine more powerful and larger (80-100 km circumference) than the LHC. It would still be based near Geneva and would use the LHC as its injector. Preparatory studies are ongoing and a conceptual design study is likely to be ready by 2017. For further details see fcc.web.cern.ch
[4] See, for example, the proposed Circular Electron Positron Collider programme (see cepc.ihep.ac.cn)
[5] Giudice (2010); Baggott (2012); Maiani (2012).
[6] For further details on the notion of the non-use value including the concept of bequest value, option value and quasi-option value see Brookshire *et al.* (1983), Johansson (1983), Bateman *et al.* (2002), Boardman *et al.* (2001).
[7] Pagiola (1996) clarifies the notion of option value, quasi-option value and existence value in the context of cultural economics.

The existence value of RIs may justify, to a certain extent, the governments' support to basic research, but little is known about the intensity of social preferences in this area. How much are people willing to pay for science? What are the rationales behind this preference? This paper is the first attempt to answer these questions by using a contingent valuation (CV) experiment.[8] Specifically, the objective of this research is twofold:

1. to grab those factors affecting individual preferences (socio-economic and cultural characteristics) for the LHC research; that is for a scientific project providing discovery as a pure public good;
2. to quantitatively estimate the non-use value of the LHC by assessing how much the general public is willing to pay for its discovery potential.

Hence, this study adds new insights to the existing literature on the non-use value of public goods by providing empirical evidence in the context of fundamental research. Our findings can be used in Cost-Benefit Analysis (CBA) of large-scale RIs, since we attach a monetary value to individual preferences for science by using standard methods set out in welfare economics. Indeed, the main benefits stemming from RIs such as the creation of knowledge outputs, technological externalities, human capital accumulation, cultural impact of the outreach, and service provision only capture the use value (or to a certain extent the option value) of these assets. In order to estimate their total economic value, the benefits related to the non-use value should be also considered (Florio and Sirtori, 2015; Florio *et al.* 2016b). Moreover, as stated by Johansson and Kriström (2015, p. 24): "If the project being evaluated affects non-use values, this should be reflected in the cost-benefit analysis". Thus, it seems interesting to examine the WTP for scientific discovery, even when, as in the case of the Higgs boson, nobody knows what its use might be.

The paper is structured as follows. Section 2 briefly describes the LHC as a provider of scientific discovery. In Section 3 we discuss the theoretical framework and the contingent valuation survey. Data and descriptive statistics are presented in Section 4. Section 5 examines the determinants of the empirical WTP by applying both binomial and polychotomous logistic models; afterwards the estimation of the individual WTP is provided as a dichotomous choice CV. Section 6 concludes.

## 2. The description of the Large Hadron Collider

The Large Hadron Collider (LHC) is the world's largest and most powerful particle accelerator, built on behalf of the European Organisation for Nuclear Research (CERN) from 1994 to 2008. It is physically located near Geneva, in a 27-kilometre tunnel, one hundred meters underground, beneath the border between France and Switzerland.

The LHC hurls beams of protons and ions at a velocity approaching the speed of light, steered by means of superconducting magnets along with a number of accelerating devices. The high-energy particle beams run in opposite directions in separate pipes and the LHC causes the beams to collide with each other at four locations around the accelerator, each corresponding at the positions of four particle detectors: ATLAS, CMS, ALICE and LHCb. Here, the resulting events caused by the collision are recorded. Collisions are examined to find answers to many issues left unsolved by the Standard Model of particles and forces[9] such as the origin of particles' mass, a coherent explanation of the interactions between the fundamental forces of the universe and the phenomena responsible for dark matter. The LHC should also help to investigate some issues related to the share of matter and anti-matter in the universe.

---

[8] As far as we are aware, Florio *et al.* (2016a) is the unique study that attempted to calculate the willingness to pay (WTP) for the LCH by using a cost-benefit analysis approach.

[9] To further details on the main goals of the LHC and on a non-physicists understandable version of Standard Model of particles and forces see the LHC guide available at http://cds.cern.ch/record/1165534/files/CERN-Brochure-2009-003-Eng.pdf

On 4 July 2012, the ATLAS and CMS experiments announced the discovery of a new particle consistent with the Higgs boson predicted by the Standard Model, adding, in this way, a new piece to our knowledge of nature.[10] In the next years, investigations of the properties of the Higgs boson as well as the possible discovery of new particles are expected to shed light on the current theory of fundamental interactions and on the puzzle about the origin of the universe.

Since its construction phase, the LHC has attracted great interest from the general public. From 2004 (when the LHC was opened to visitors) to 2013, 418,200 people visited the LHC, reaching a peak of about 100,000 visits per year in the aftermath of the announcement of the Higgs boson discovery.[11] The travelling exhibitions related to the LHC have attracted 344,000 visitors worldwide to 2013. In the same period, the users of LHC-related social media (YouTube; Twitter; Facebook; Google+) amounted to around 2,010,000; while the number of CERN-LHC website visits was more than 37 million. These figures are expected to further increase in the near future (Florio *et al.*, 2016a).

The focus of our paper is specifically about the interplay between basic science and its perception by the general public, which is apparently interested in it, and in fact supports it indirectly through taxation, considered that the CERN is entirely funded by transfers by its Member States.

## 3. Conceptual framework

*3.1 The contingent valuation approach*

The approach used in this study to estimate WTP for the LHC is the contingent valuation (CV) method. Since the estimation of the existence value of RIs is still in its infancy, it was decided to recur to a well-established framework such as the CV method rather than adopting more recent techniques as, for example, stated choice experiments, which would pose a number of problems in our context.[12] This refinement is left to further research.

CV is the most common and more reliable approach to estimate the non-use values, and in particular, the existence value of public goods (Davis, 1963; Mitchell and Carson, 1989; Carson and Groves, 2007). It is a survey-based method that aims at investigating how people would behave in a hypothetical situation. A standard reference for CV studies are the guidelines provided by Arrow *et al.* (1993), which were endorsed by an authoritative panel of experts convened by the National Oceanic and Atmospheric Administration (NOAA). These guidelines have then been extensively refined, adapted and implemented to estimate the non-market value of public goods, particularly environmental, health and cultural goods (Carson *et al.*, 1996; Hansen, 1997; Whitehead and Hoban, 1999; Thompson *et al.*, 2002; Venkatachalam, 2003; Polome *et al.*, 2006; Carson and Groves, 2007; Carson, 2012).

In order to elicit the respondents WTP for a given good, the NOAA guidelines suggest to use a referendum-like approach[13] according to which, people should be asked to state only '*yes*' or '*not*' to the proposed bid. Hanemann (1984) and Carson (1985) have stressed the importance of follow−up questions addressed to estimate the maximum WTP value. The bid level offered in the follow−up question should be greater than that offered in the initial payment offer if the answer to the initial payment question is '*yes*', otherwise the follow-up procedure is stopped. Although this approach is statistically more efficient than referendum approach, Alberini *et al.*, (1997) found that the average WTP estimated after the follow−up approach can be lower than that implied by the responses to the initial payment question. A possible explanation is that some respondents may treat the suggested bid as a signal for the quality of the good and/or might erroneously believe that the program to be valued in the follow−up is different from the initial one. On the other side, single referendum elicitation format is highly vulnerable to anchoring effects (Green *et al.*, 1998).

---

[10] The Higgs mechanism may shed light on the mass of particles, and may explain why some particles are very heavy while others have no mass at all. According to the Higgs mechanism, the whole of space is filled with a Higgs field, and the way through which particles interact with this field, determines their specific masses. The Higgs boson is one of the new particles predicted by the Higgs mechanism.
[11] Our elaborations on Florio *et al.* (2016a) data.
[12] More recent references on how to efficiently design non-market valuation surveys are, for example, DeShazo and Fermo (2002) and Rose and Scarpa (2008).
[13] The referendum-like approach is also known as single-bounded dichotomous choice approach.

*3.2 The CV questionnaire*

Drawing from the methodological insights of this literature, we carried out a CV survey to achieve two goals. First, to detect the explanatory variables potentially affecting individuals' WTP for the LHC. Second, to estimate the expected individual WTP for the LHC as a provider of scientific discovery. The questionnaire was designed to be consistent, as far as possible, with the NOAA panel guidelines by Arrow *et al.*, (1993); while some modifications were applied to take into account the peculiarities of the good under evaluation. A pilot survey was conducted with students at the University of Milan in order to calibrate the structure, the number of questions and the duration of the interview as well as to verify the questionnaire was readable and clear so as to reduce the rate of rejection from respondents.[14]

The questionnaire was structured along three sections. Section A addressed to investigate background knowledge and broad awareness of respondents about RIs. Open-ended and five-point Likert scale questions were added to binary-choice questions to further detect the interviewees' preferences and interest towards research. A brief description of the LHC was provided to interviewees: a shortened version of the Wikipedia entry "Large Hadron Collider", including five photos. Section B focused on LHC and included questions to elicit the WTP (see discussion below). Section C contained personal information and, specifically, questions about socio-economic characteristics affecting individual preferences. In particular, it inquired on age, sex, country of residence, university studies, income, and household composition.

In section B, the WTP was inquired in two ways. First, respondents were asked about their willingness to offer a single lump-sum payment amounting to EUR 30. They had three possible alternative answers: '*yes*', '*no*' and '*do not know*'. The amount of EUR 30 comes from Florio *et al.* (2016b), who has carried out a meta-analysis of CV studies worldwide on the existence value of public goods, particularly environmental, health and cultural goods. Second, the WTP was asked in the form of an annual fixed contribution for a period of 30 years. In particular, the bids of EUR 0, EUR 0.5, EUR 1 and EUR 2[15] per year were proposed, by stemming from the following facts. At a general public conference on the LHC at CERN in 2008, Dr. Fabiola Gianotti, General-Director of CERN since 2016, stated that CERN costs to the tax-payers "a cup of coffee per year".[16] Since then, this citation has been repeatedly used by LHC stakeholders.[17] Thus, the offered options were centered around the "one cup of coffee per year" benchmark and respectively one half or twice that reference point. The 30 years' time-span suggested in our survey is the period that goes from the approval of the LHC project's budget in 1996 to the LHC planned decommissioning in 2025. The same time period was used in the CBA of the LHC by Florio *et al.* (2016a). Thus the 30 Euro lump-sum is also roughly equivalent to 'one cup of coffee over 30 years'.

Some CV studies test whether the mean WTP changes according to the manner the WTP question is asked, that is periodic payment versus lump-sum payment (see for example Echevveria *et al.*, 1995). In the present study, we are not interested in comparing the expected WTP coming from the two ways of asking the CV question, instead we would like to know whether the determinants of WTP are different according to the way the WTP is asked and in this case to know what motivates such difference. Hence, we proposed respondents to fill in both the CV questions and in both cases they have been reminded that their willingness to pay for the LHC would reduce their budget for other goods.

Respondents were also asked to explain their choice by filling an open-ended question as a proof of his/her sincerity and to identify "protest bids"; i.e. investigate the reason why people are not be willing to pay for the good (Dubgaard, 1996). Some authors (Loomis *et al.*, 1996; Hansen, 1997) understand protest bids as a rejection of the CV study approach itself or as an expression of the political ideology of the respondent. If protest bids were identified, they will be sifted out from the analysis.

---

[14] The pilot test involved a total of 61 students. More details on the survey, the questionnaire and on the results are presented in Catalano *et al.* (2014). The questionnaire was originally submitted in Italian during the pilot test and then translated in English, French and Spanish. The full questionnaire is enclosed in Annex I.
[15] See section 4 for further details.
[16] https://indico.cern.ch/event/416935/session/4/contribution/5/attachments/859149/1201296/esof2008.pdf
[17] http://vmsstreamer1.fnal.gov/Lectures/LectureSeries/presentations/110415Meddahi.pdf

We chose to add a '*do not know*' option to '*yes*' and '*no*' options when asking about the lump-sum WTP. This structure seems particular convenient when people are willing to pay for something beneficial for the human-kind, as the fundamental research is supposed to be, but people do not know what precisely they should pay for. Thus, the benefit of offering a middle response in addition to the referendum-like options '*yes*' and '*no*' is that uncertain survey respondents are not forced to construct a willingness to pay when answering a dichotomous choice question. The cost is a reduction in sample size and econometric efficiency if the '*do not know*' answers are dropped from the empirical analysis (Hansen, 1997; Groothuis and Whitehead, 2002).

Face-to-face interviews were conducted by professional interviewers in order to minimize interview bias which arises when the interviewer accidentally lead respondent in a particular direction when answering the questionnaire. Furthermore, respondents' strategic behavior was tackled by ensuring the anonymity of the questionnaire so as to reduce suspicions related to highly sensible information (Bohm, 1972; Arrow *et al.*, 1993).[18]

*3.3 Investigating the determinants of WTP*

Our first research question was to examine the explanatory variables affecting individuals' WTP. To this end, we adopted two approaches: a zero-inflated ordered logistic model was used when the dependent variable was the WTP expressed in the form of an annual fixed contribution while a standard multinomial logistic model was applied when the WTP was asked as a single lump-sum payment.

In the first case, the WTP is a discrete ordered variable, including the zero value. As demonstrated by Harris and Zhao (2007), traditional ordered logit models have limited capacity in explaining zero observations. This suggests to use a zero-inflated ordered logit model by applying a double combination of a split logit model and an ordered logit model. Assume we observe $N$ independent observations where the dependent variable of interest, $WTP_i$ ($i = 0, 1, ..., N$) is a count variable that displays a large proportion of zeros. Each individual belongs to one of two groups. The first group includes individuals who chose the zero option (they were not willing to pay) at the time the survey was conducted. The second group includes respondents for whom the WTP was positive. This suggests a two-stage approach. In the first stage we examine the binary decision to be willing to pay or not to pay. In the second stage, we investigate the probability of falling in one of the bid categories conditional on the decision to be willing to pay (Harris and Zhao, 2007; Kasteridis *et al.* 2010).[19]

The binary decision to pay is modelled with a logit model. Let the underlying response variable $p_i^*$ be defined by the latent regression:

$$p_i^* = X_{1i}\gamma + u_i \qquad (1)$$

where the latent variable $p_i^*$ measures the difference in utility derived by individual $i$ from being willing to pay and not being so, $X_1$ is a vector of exogenous variables affecting individuals' preferences, $\gamma$ is a vector containing all the parameters in the model and $u$ is the error term. The observed binary variable for being willing to pay ($P_i$) relates to the latent variable ($p_i^*$) such that:

$$P_i = \begin{cases} 1 & if\ p_i^* > 0 \\ 0 & otherwise \end{cases} \quad (2)$$

---

[18] Strategic behaviour occurs when a systematic error is introduced into the sampling, when respondents select one answer over others in order to not to reveal their true opinion/position. A well-known case is that of perceived government-supported surveys leading people to skip highly-sensible information like income.

[19] In principle we may estimate an ordered choice model on the entire sample, including those individuals with zero WTP. As explained by Harris and Zhao (2007), this strategy could invalidate our results because the presence of many zeros in our response variable may indicate two distinct data generating processes and WTP behaviours. If so, a single latent equation model, as the ordered logit model, would not allow for the differentiation between these two aspects. Differently, we may estimate a truncated ordered choice model by throwing out zero observations incurring, in this case, in the sample selection problem (Greene, 2012; Verbeek, 2008). This trade-off between alternative options lead us to implement a two-stage approach.

From equation (1) and (2) the probability of paying is

$$\Pr(P_i = 1) = \Lambda(X_{1i}\gamma) \quad (3)$$

where $\Lambda(.)$ is the logistic cumulative distribution function (cfd) of $u_i$. The observed values of $P_i$ are the realisations of a binomial variable with probabilities given by equation (3) conditional on $X_{1i}$. Thus the likelihood function is given by:

$$L(\gamma|P_i, X_{1i}) = \prod_{P_i=0}[1 - \Pr(P_i = 1)] \prod_{P_i=1} \Pr(P_i = 1) \quad (4)$$

Once modeled the decision to be willing to pay or not to pay, we focus on the level of the bid chosen. Let $WTP_i$ be our dependent variable measuring the level of the proposed bids and takes on integer values from 0 to J. The variable $WTP_i$ takes the value of zero for those respondents who chose the zero option ($P_i = 0$). Nonzero values can only be observed conditional on $P_i = 1$. The joint likelihood of observing the entire sample is:

$$L(\theta|WTP_i) = \prod_{P_i=0}[1 - \Pr(P_i = 1)] \prod_{P_i=1} \Pr(P_i = 1) \Pr(WTP_i|P_i = 1) \quad (5)$$

where $\theta$ is a vector containing all the parameters in the model and the products are taken over sample observations satisfying ($P_i = 0$) and ($P_i = 1$). The conditional density $\Pr(WTP_i|P_i = 1)$ can be handled by an ordered logit model (Greene 2012; McElvey and Zavoina, 1975). In our model each individual reveals the strength of her preferences with respect to the level of bid chosen. Although the preferences will probably vary continuously in the space of individual utility, the expression of those preferences is given in a discrete outcome on a scale with a limited number of choices. As a result, the model is constructed around a latent regression of the following form:

$$WTP_i^* = X_{2i}\beta + \varepsilon_i \quad (6)$$

where $WTP_i^*$ can be interpreted as the propensity to be willing to pay at different bids, $X_2$ is a vector of exogenous variables aiming at explaining the level of bid and $\varepsilon_i$ is the random disturbance term that follows a logistic distribution. The variable $WTP_i$ relates to the latent variable ($WTP_i^*$) according to the rule:

$$\begin{aligned} WTP_i &= 1 \quad if \quad\quad\ WTP_i^* \le \tau_1 \\ WTP_i &= j \quad if \quad \tau_{j-1} < WTP_i^* \le \tau_j \quad j = 2, \dots, J-1 \\ WTP_i &= J \quad if \quad \tau_{J-1} < WTP_i^* < \infty \end{aligned}$$

where $\tau_1 \le \tau_2 \le \cdots \le \tau_{J-1}$ are unknown thresholds (cut-points) to be estimated.
The conditional distribution of $WTP_i$ given $P_i = 1$ and $X_{2i}$ and the likelihood function of this sub-sample are given respectively by equations (7) and (8)

$$\Pr(WTP_i = j|P_i = 1) = \Lambda(\tau_j - X_{2i}\beta) - \Lambda(\tau_{j-1} - X_{2i}\beta) \quad (7)$$

$$L(.\mid WTP_i, X_{2i}) = \prod_{j=1}^{J}[\Lambda(\tau_j - X_{2i}\beta) - \Lambda(\tau_{j-1} - X_{2i}\beta)]^{WTP_{ij}} \tag{8}$$

where $WTP_{ij}$ is an indicator variable equals to 1 if $WTP_i$ falls in the j-th category and 0 otherwise.

Ideally, we should estimate the model presented in equation (5). From a statistical perspective, it describes the joint distribution of the random variables $WTP_i$ and $P_i$ conditional on the explanatory variables contained in $X_{1i}$ and $X_{2i}$ with the variance-covariance structure of the bivariate distribution of the error term $\xi \equiv [u_i \; \varepsilon_i]'$ defined by the following matrix:

$$V(\xi) = \begin{bmatrix} \sigma_\varepsilon^2 & \sigma_{\varepsilon u} \\ \sigma_{\varepsilon u} & \sigma_u^2 \end{bmatrix} \tag{9}$$

In this model, $X_{1i}$ includes the variables that determine the decision process to be willing to pay or not to pay (the participation equation, 3) and $X_{2i}$ those that influence the level of the bid chosen (the level equation, 7). In general, $X_{1i}$ is an informative set contained in $X_{2i}$. Unfortunately, our dataset does not allow us to provide variables that would affect the participation equation but not the level equation or vice versa, so we have that $X_{1i} = X_{2i}$. Being so, we impose the covariance between $u_i$ and $\varepsilon_i$ to be zero. This allows us to estimate equation (3) and equation (7) separately; specifically we are going to estimate the equation (3) by using the whole sample and the equation (7) by using an ordered logit model over the sub-sample in which we observe $P_i = 1$.[20]

The determinants of WTP when expressed in the form of a single lump-sum payment are investigated by estimating a standard multinomial (MNL) logit model (Maddala, 1994). For sake of brevity and clarity, we don't treat the theory behind the MNL procedure (for details see Maddala, 1994; Greene, 2012).

### 3.4 Estimating the WTP

The reference model to estimate the existence value of environmental and cultural goods is the utility difference model developed by Hanneman (1984). We use this model to provide an answer to our second research question, which is the average contribution that the respondents would be willing to pay for LHC. Let's assume that the dependent variable of interest, $S_i$ ($i = 0,1$) is a binary variable. $S_i = 0$ identifies individuals who would not be willing to pay for the good being evaluated; in contrast, $S_i = 1$ identifies people willing to pay the bid proposed by the interviewer. Each individual has an indirect utility function of the form $V(M; Y_i; Z_i)$ where $Y_i$ is income, $Z_i$ is vector of exogenous variables affecting individuals' preferences (in our case socio-economic characteristics and attitude towards research) and $M$ is a binary variable describing the state of the world with or without the good under evaluation.

When interviewed, the respondent has two options: (a) to answer '*no*' and face the state of the world in absence of the good ($M = 0$) and keep all of his/her income ($Y_i$); (b) to choose '*yes*' and thus having his/her income reduced by the bid ($A$) but the good available for the future ($M = 1$). An individual will respond '*yes*' if and only if his/her utility under option (b) is greater than or equal to that under option (a): $\delta V_i^* = V(1; Y_i - A; Z_i) - V(0; Y_i; Z_i) + v_i \geq 0$.

Empirically, the probability that the individual accepts the offer ($A$) is approximated with a binomial model given by

$$\Pr(S_i = 1) = \Lambda(\delta V_i^*) = \Lambda(\alpha + A\beta_1 + Y_i\beta_2 + Z_i\beta_3) \tag{10}$$

---

[20] This is a limitation of the study. However, several papers in different fields use this approach. See for example Geda *et al.* (2001), Lera-Lopez and Rapun-Garate (2007), Kasteridis *et al.* (2010).

where the latent variable $\delta V_i^*$ measures the difference in utility, $\Lambda(.)$ is the logistic cdf of the error term $v$ and $\alpha, \beta_1, \beta_2, \beta_3$ are the parameters of the model to be estimated, where $\beta_1 \leq 0$ and $\beta_2 > 0$ are expected.

Once equation (10) is estimated, the expected value of WTP is obtained by numerical integration. As argued by Duffield and Patterson (1991) there are three methods to compute the value of WTP. The first is to compute the WTP by integrating equation (10) from $-\infty$ to $+\infty$ obtaining the so called overall mean WTP. Since the WTP is nonnegative, this method is not appropriate. Thus the two alternative approaches are to compute the expected value of the WTP integrating from 0 to $+\infty$ or the truncated mean WTP integrating from 0 to maximum bid ($A$). The authors suggest that the truncated mean WTP is the most appropriate method because satisfies theoretical constraints (the upper limit of the WTP is not infinity but something less than income), is statistical efficient in the sense that reduces the influence of the upper tail of the empirical distribution of WTP and satisfies the aggregation criteria. By using this method the value of the maximum bid ($A$) has to be assigned to all recorded WTP above ($A$). Thus:

$$E(WTP) = \int_0^{MAX\ A} \Lambda\left(\delta V_i^*(A)\right) \delta A =$$
$$= \int_0^{MAX\ A} \left[+ \exp\left(-\left(\hat{\alpha} + A\hat{\beta}_1 + Y_i\hat{\beta}_2 + X_i\hat{\beta}_3\right)\right)\right]^{-1} \delta A \quad (11)$$
$$= \int_0^{MAX\ A} \left[1 + \exp\left(\hat{\alpha}^* + A\hat{\beta}_1\right)\right]^{-1} \delta A$$

where $\hat{\alpha}^*$ is the estimated adjusted intercept wich was added by the socio-economic characteristics and reasearch-attitude factors to the original constant $\hat{\alpha}$.

In what follows, the truncated mean WTP is obtained by making use only for the lump-sum payment of EUR 30. For the expected WTP estimated by using the data on the annual payments see Florio *et al.* (2016a).

## 4. Data and descriptive statistics

The CV survey was conducted between June 2014 and March 2015. 1022 valid questionnaires were filled in by students coming from five European universities located in four countries: University of Milan (Italy), University of Exeter (UK), University Paris 7- Denis Diderot and Sciences Po University (France), University of A Coruña (Spain).

The descriptive statistics of the variables related to the profile of students in terms of social features and respective codes are presented in Table 1. The sample comprises 420 (41%) students from Italy, and about 200 (20%) from Spain, France and the UK each. They are enrolled in more than 30 different university degrees with an overall balance between three main fields of education: 39% of respondents are enrolled in humanistic related faculties[21], 25% in social sciences[22] and 34% in scientific degrees[23]. 578 (57%) respondents are female and 86% (857 students) is aged between 19 and 25 years, while the remaining share is more than 26 years old. Most of students belong to a family from 3 to 5 members (774 corresponding to 75.7% of the sample), 174 (17%) to a family with 1 or 2 members and only a tiny share (7.2%) to a family with more than 5 people.

---

[21] These include law, foreign languages, international relations, literature, philosophy, history, geography, cultural assets, communication and media, theology, cryptography, musicology.
[22] Including economics, finance, marketing and management, political sciences, sociology, semiology, anthropology, humanitarian sciences, sport sciences, urban studies, education.
23 Such as medicine, pharmacy, chemistry, biology, mathematics, physics, engineering, architecture, mechanics, ICT.

**Table 1. Descriptive statistics of social characteristics**

| Variable | Code | Number | Percent |
|---|---|---|---|
| C0.1 Country | 1= Italy | 420 | 41.1 |
|  | 2= Spain | 202 | 19.8 |
|  | 3= France | 200 | 19.6 |
|  | 4= UK | 200 | 19.6 |
| C0.2 Education [a] | 1= Humanities | 398 | 38.9 |
|  | 2= Social sciences | 257 | 25.1 |
|  | 3= Scientific | 352 | 34.4 |
| C1 Age | 1= 19-25 | 875 | 85.6 |
|  | 2= 26-30 | 94 | 9.2 |
|  | 3= 31-35 | 33 | 3.2 |
|  | 4= > 35 | 20 | 2.0 |
| C2 Gender | 0=Male | 444 | 43.4 |
|  | 1= Female | 578 | 56.6 |
| C8 Household Composition | 1= 1-2 | 174 | 17.0 |
|  | 2= 3-5 | 774 | 75.7 |
|  | 3= > 5 | 74 | 7.2 |

Question C0.1: Country of residence, Question C0.2: Faculty, Question C1: Age, Question C2: Sex, Question C8: Household composition (including parents, brothers/sisters). [a] 15 observations are missing

Regarding income, respondents were inquired both on the availability of a personal income and on the amount of the family income. The joint distribution of income variables is reported in Table 2. Numbers in parentheses denote the distribution function of the single variable. Only 304 (30%) students have an own income. The largest share (70%) is financially supported by his/her family. Most of households fall in the income category ranging from EUR 1,000 to 3,000 per month (478 respondents, representing 47% of the sample), followed by a 23% share with monthly income ranging from EUR 3,000 to 5,000. 19% and 11% of respondents fall in the lowest (less than EUR 1,000) and highest (more than EUR 5,000) family-income categories respectively. The Spearman's correlation coefficient between the two variables is negative (-0.03) meaning that students who earn an own income come from families with a low-medium income. Table 2 shows that 65.2% of respondents with a personal income belong to families with less than EUR 3,000 per month. In the following analysis, it was decided to use the family income (Question C6) as independent variable rather than the availability of own income (Question C7). This approach is based on the fact that only 30% of respondents earn their own income; therefore it is very likely that their decision-making process about WTP is highly influenced by the family budget constraints.

**Table 2. Descriptive statistics of income variables**

|  |  | C7 Respondent's availability income | |
|---|---|---|---|
|  |  | 0=No | 1=Yes |
| C6 Family monthly income (EUR) [a] |  | 718 (70.3%) | 304 (29.7%) |
| 1= < 1,000 | 189 (18.5%) | 15.9% | 25.4% |
| 2= 1,000 - 3,000 | 478 (46.8%) | 50.3% | 39.8% |
| 3= 3,000 - 5,000 | 231 (22.6%) | 23.1% | 22.4% |
| 4= > 5,000 | 113 (11.1%) | 10.7% | 12.4% |

Question C6: In which of the following brackets does your family monthly net income fall?
Question C7: Do you have your own personal income? [a] 11 observations are missing

Previous knowledge and awareness of respondents about RIs as well as their attitude towards scientific discoveries have been investigated before going to the specific case of LHC. Results are presented in Table 3. Out of 1022 respondents, 554 (54.2%) are aware of what exactly a RI is and 480, representing the 53% of the whole sample, associate it with a particle accelerator when asked to identify a RI amongst some alternatives.[24] 845 (83%) interviewees stated that they have an interest for scientific discoveries, and more in general for the research activity and 85% recognises that funding RIs is at least important. The LHC is known by 535 (52.3%) interviewees. Their source of information includes mainly internet, magazines and TV (62.4%). 117 (21.9%) students declared that they have heard about LHC at university or related activities such as seminars and meetings while 83 (15.7%) by friends. Higgs Boson is known by 620 (60.7%) respondents and 97 (9.3%) had already visited CERN.

**Table 3. Interest in research and knowledge of LHC**

| Variable | Code | Number | Percent |
| --- | --- | --- | --- |
| A1 Knowing what a RI is | 0= No | 468 | 45.8 |
|  | 1= Yes | 554 | 54.2 |
| A2 Particle accelerator | 1= Particle accelerator | 480 | 53.0 |
|  | 0= Other [c] | 542 | 47.0 |
| A4 Interest in research | 0= No | 177 | 17.3 |
|  | 1= Yes | 845 | 82.7 |
| A6 Importance of funding RI | 1= Useless | 4 | 0.4 |
|  | 2= Insignificant | 13 | 1.3 |
|  | 3= Important Enough | 142 | 13.9 |
|  | 4= Important | 473 | 46.3 |
|  | 5= Fundamental | 390 | 38.2 |
| B1 Having heard about LHC | 0= No | 487 | 47.7 |
|  | 1= Yes | 535 | 52.3 |
| B2 Source of information about LHC | University | 117 | 21.9 |
|  | TV | 119 | 22.3 |
|  | Magazines | 86 | 16.1 |
|  | Internet | 130 | 24.0 |
|  | Friends | 83 | 15.7 |
| B3 Having heard about Higgs Boson | 0= No | 402 | 39.3 |
|  | 1= Yes | 620 | 60.7 |
| B5 Having visited the CERN | 0= No | 927 | 90.7 |
|  | 1= Yes | 97 | 9.3 |

Question A1: Do you know what a research infrastructure is?
Question A2: In your opinion, which of the following is a research infrastructure? telescope; instrument of data collection and archive; data elaboration software; particle accelerator; library; computer; astronomical observatory; planetarium.
Question A4: Are you interested in scientific discoveries and in research activities in general?
Question A6: how do you rate the importance of funding research infrastructures?
Question B1: Did you hear about the LHC before this questionnaire?
Question B2: If yes, please indicate your source of information.
Question B3: Did you ever hear of "Higgs Boson"?
Question B5: Have you ever been to the CERN?
[a] Elaborations on B1=Yes

Table 4 presents the descriptive statistics and the joint distributions of the questions related to WTP. Again, numbers in parentheses denote the distribution function of the single variable. Respondents stated their willingness to pay for the LHC in the following way. First, a general question was submitted to detect the students' willingness to pay for the research activity at the LHC, without mentioning any bid (Question B8). 496 (48%) respondents declared that they do not know whether they are willing or not to contribute to the activity of LHC, 335 (33%) explicitly declared they would not be willing to financially support the LHC, while 191 (19%) respondents were willing to financially support the LHC (Table 4, Panel A)

---
[24] Multiple choices were allowed to Question A2 of the questionnaire amongst the followings: telescope, instrument of data collection and archive, data elaboration software, particle accelerator, library, computer, astronomical observatory, planetarium.

Afterwards, two questions integrating, respectively, two different payment systems, were submitted. The first question (B10) was: "*By 2015, would you be willing to offer an economic contribution equals to EUR 30 turning down other personal expenses?*" with the offered answers being '*yes*', '*no*' and '*do not know*'. The second question (B12) was: "*If someone asks you to give an economic contribution to the LHC by means of an annual tax over a period of 30 years, would you be willing to pay an annual amount equal to: EUR 0, EUR 0.5, EUR 1 and EUR 2*". The survey reveals that 147 interviewees (14%) are willing to pay EUR 30 una tantum, 500 (49%) would not pay as much, and 375 (37%) would not know (Table 4, Panel A).

When looking at the second question, the share of respondents who would be willing to contribute EUR 0.5, EUR 1 and EUR 2 are respectively 83 (8%), 229 (22%) and 438 (43%). The remaining 27% (274 students) would be willing to pay EUR 0 (Table 4, Panel B). The distribution of the WTP responses is also presented in Figure 1 and 2, where the answers categories are plotted on the *X* axis, while the number of respondents is plotted on the *Y* axis.

In both WTP scenarios, five protest bids were identified. Three respondents said '*no*' in the lump-sum scenario and chose EUR 0 in annual payment scenario because national governments should be in charge of funding RIs; while two respondents were against the allocation of resources for the LHC in time of crisis. Protest bids represent a negligible share (2%) and they have no effects on our findings. The most quoted motivations behind not to be willing to pay were the non-affordability of the bid offered, the low interest in scientific discoveries and the lack of sufficient information about LHC. These responses represent valid reasons for not being willing to pay.

Interesting findings are showed in Table 4. Firstly, it is worth noting that the share of respondents who would be willing to contribute at least EUR 0.5 to the LHC via an annual tax is much higher (73%) than that declared in the lump sum question (14%). Among this 73%, about 60% would be willing to pay an annual contribution of EUR 2, that is the highest offered option. These figures confirm that people react differently to different payment proposals in line with many CV studies (Echevveria *et al.*, 1995; Green *et al.*, 1998). In the next section, we investigate in depth the individuals' characteristics affecting such making decision process.

Secondly, Table 4 suggests an empirical strategy to estimate the expected WTP for the LHC. In order to apply the Hanneman's model, we exploit the variation in the data coming from the joint distribution of the questions B8 and B10. Specifically, our strategy consists of identifying those respondents whose maximum WTP is EUR 30 or even more. Panel A shows that there are 500 students who answered '*no*' to the follow-up question (B10) and 48 respondents with answers of '*yes-do not know*' type. For these students, it is very likely that their WTP falls between EUR 0 and an (unknown) amount smaller than EUR 30. In contrast, the group of students (147) who answered '*yes*' to the question (B10) are willing to pay at least EUR 30. At this point, we construct a dummy variable (BID) that takes on the value 1 for these respondents and 0 otherwise. By applying this strategy, we also decided to throw 282 answers with the '*do not know-do not know*' combination away and estimate the expected WTP by using a restricted sample with only '*yes*' and '*no*' records. This choice allows us to elicit only the stronger opinions on the issues; the cost is the loss of efficiency in the econometric analysis (Groothius and Whitehead, 2002). Indeed, these restrictions reduce our sample from 1,022 to 740 individuals.

Table 5 analyses the overall correlation between the variables presented above. As the variables are expressed in ordinal intervals, we use the Spearman's rank correlation matrix. The coefficients show that family income (C6) has a significant positive correlation with the WTP when expressed as a lump-sum payment (B10). In contrast, it shows a non-significant association with the WTP when expressed in the form of an annual contribution (B12). Moreover, the greater the family income, the greater the family composition. The variable BID is negatively associated with both WTP variables, and in particular with the lump-sum payment. There is also a positive and statistically significant relationship between education and willingness to pay (B12) meaning that students enrolled in scientific faculties are willing to contribute more for the LHC.

As expected the variables expressing interest in research (A4 and A6) and the variables related to the LHC, the Higgs boson and the CERN (B1, B3 and B5) positively correlate with WTP. It is worth noting that the latter variables are also strongly correlated each other. Hence, to avoid significant multicollinearity we leave the variable B5 out in the econometric exercise.[25]

**Table 4. Descriptive statistics of the WTP questions**

| PANEL A | (B10) By 2015, would you be willing to offer an economic contribution equal to 30 euros (lump sum) turning down other personal expenses? | | | |
|---|---|---|---|---|
| (B8) Would you be willing to provide an economic contribution to fund the research activity of LHC? | No | Yes | I do not know | Total |
| No | 287 | 3 | 45 | 335 (33%) |
| Yes | 54 | 89 | 48 | 191 (19%) |
| I do not know | 159 | 55 | 282 | 496 (48%) |
| Total | 500 (49%) | 147 (14%) | 375 (37%) | 1,022 (100%) |

| PANEL B | (B12) If someone asks you to give an economic contribution to the LHC by means of an annual tax over a period of 30 years, would you be willing to pay an annual amount equal to: | | | | |
|---|---|---|---|---|---|
| | EUR 0 | EUR 0.5 | EUR 1 | EUR 2 | Total |
| No | 179 | 28 | 48 | 80 | 335 |
| Yes | 6 | 6 | 47 | 132 | 191 |
| I do not know | 87 | 49 | 134 | 226 | 496 |
| Total | 274 (27%) | 83 (8%) | 229 (22%) | 438 (43%) | 1,022 (100%) |

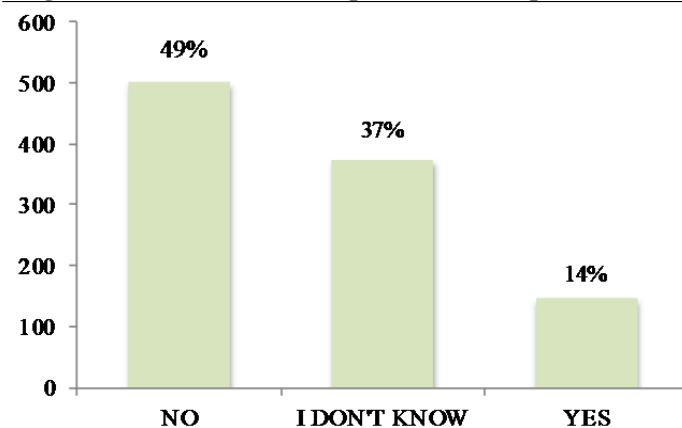

Figure 1. Distribution of lump-sum WTP responses

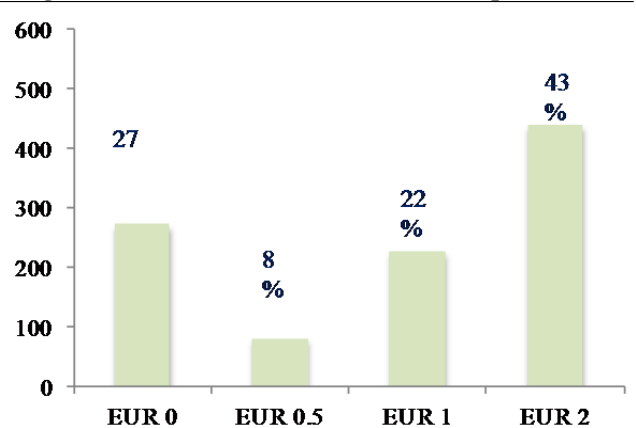

Figure 2. Distribution of annual WTP responses

---

[25] The Spearman's correlation coefficients between the variable "B5 Having visited CERN" and "B1 Having heard about LHC" and "B3 Having heard about Higgs boson" are 0.50 and 0.7 respectively.

**Table 5. Spearman's Rank Correlations**

|  | B10 | B12 | BID | C0 | C1 | C2 | C6 | C8 | A1 | A2 | A4 | A6 | B1 | B3 | B5 |
|---|---|---|---|---|---|---|---|---|---|---|---|---|---|---|---|
| B10 WTP Fixed contribution | 1 | | | | | | | | | | | | | | |
| B12 WTP Annual payment | 0.26** | 1 | | | | | | | | | | | | | |
| BID | -0.55* | -0.09* | 1 | | | | | | | | | | | | |
| C0 Education | -0.01 | 0.11** | -0.04 | 1 | | | | | | | | | | | |
| C1 Age | -0.03 | -0.10** | -0.05 | -0.01 | 1 | | | | | | | | | | |
| C2 Gender | 0.05 | 0.01 | -0.02 | -0.02 | -0.05 | 1 | | | | | | | | | |
| C6 Family monthly income | 0.09** | 0.02 | 0.02 | -0.06** | -0.02 | -0.05 | 1 | | | | | | | | |
| C8 Household composition | 0.03 | 0.01 | 0.02 | 0.05 | -0.01 | 0.03 | 0.07** | 1 | | | | | | | |
| A1 Knowing what a RI is | 0.03 | 0.06 | 0.05 | -0.12** | -0.06** | 0.10** | 0.01 | 0.04 | 1 | | | | | | |
| A2 Particle accelerator | 0.03 | 0.07** | 0.05 | 0.14** | 0.05 | -0.18** | -0.04 | 0.01 | 0.37** | 1 | | | | | |
| A4 Interest in research | 0.10** | 0.05 | 0.10* | 0.13** | 0.10** | -0.02 | 0.04 | -0.05 | 0.07** | 0.06** | 1 | | | | |
| A6 Importance of funding RI | 0.02** | 0.15** | 0.03 | 0.06 | 0.03 | 0.00 | 0.01 | -0.08** | 0.15** | 0.13** | 0.46** | 1 | | | |
| B1 Having heard about LHC | 0.02 | 0.17** | 0.03 | 0.15** | 0.02 | -0.20** | 0.07** | -0.00 | 0.08** | 0.26** | 0.10** | 0.16** | 1 | | |
| B3 Having heard about Higgs Boson | 0.05 | 0.15** | 0.05* | 0.11** | 0.00 | -0.20** | 0.07** | 0.01 | 0.01 | 0.10** | 0.27** | 0.17** | 0.14** | 0.47** | 1 |
| B5 Having visited the CERN | 0.02 | 0.07** | 0.01 | 0.06** | 0.00 | -0.02 | 0.02 | 0.01 | -0.00 | 0.04 | 0.04 | 0.10** | 0.05 | 0.50** | 0.66** | 1 |

** Significant at 5% level

## 5. Econometric results

### 5.1 *Determinants of WTP*

We start with examining the determinants of individuals' WTP when expressed in the form of an annual fixed contribution for a period of 30 years with four options: EUR 0, EUR 0.5, EUR 1 and EUR 2. ML estimates of equation (3) and the marginal effects of each variable are presented in Table 6. In order to estimate the binomial logit model, we convert the dependent variable in a binary variable, which takes the value of 1 for they would be willing to pay and 0 otherwise. Therefore the model explains the probability of observing a positive WTP.

In column (1), we examine the impact of the socio-economic features leaving out the variables expressing attitude and interest towards research. The latter were added in column (3). In both cases we control for country-specific effects. The latter are particularly relevant within this study because the WTP of the respondents may depend on the extent to which a country contributed to the funding of the LHC or on the country's exposure to the CERN.[26] These dummies should be able to capture it the extent to which there is unobserved country heterogeneity. Average marginal effects of the two specifications are reported in columns (2) and (4) respectively.

Income enters positively and significantly into both models. For example, belonging to a family with monthly income between EUR 3,000 and 5,000 increases the probability of declaring a positive WTP by about 10 percentage points compared to a student belonging to a family with an income less than EUR 1,000 ceteris paribus (Columns 2 and 4).

The coefficients on age variables are statistically significant at either the 10% level or 5% level with negative sign. The negative sign indicates that the probability of WTP '*yes*' is likely to be higher in younger students than in older one.[27]

Surprisingly, education is not statistically significant in either equation, in contrast with most of CV studies which show that, on average and depending on the good under evaluation, a more educated population is more likely to be willing to pay for public goods (Whitehead and Blomquist, 1991; Echeverria *et al*., 1995; Thompson *et al*., 2002; Amirnejad *et al*., 2006). Accordingly, Florio *et al*. (2016a) argue that more educated people would probably be willing to pay more for the LHC than less educated people. The lack of significance of education in the participation equation, can be easily explained by noting that our survey did not target people with different educational attainments, but university students enrolled in several curricula; so the variation in education across respondents is not in terms of years spent in education but exclusively in terms of background and subject preferences. Amongst the source of information about the LHC, on-line news and TV programmes are the most quoted by the interviewees. Being so, is not an anomaly to observe no differences in the probability of observing a positive WTP between university degrees.

Finally, no significant gender and family composition differences exist on the WTP; in contrast knowing what is a particle accelerator, having heard about the LHC and the Higgs Boson and declaring that the fundamental research is important to some extent, increase the probability of a '*yes*' response.[28]

Table 7 presents the estimates of the ordered logit model conditional on the sub-sample of student who are willing to pay (Equation 7). Also reported in the table, there are the marginal effects of explanatory variables in model 2 for each category of WTP.

Column 2 shows that income is negatively associated with the level of WTP. The higher the income, the less likely is the probability of choosing EUR 2. Economic theory, structure of the question and/or the data themselves can help us to explain this apparently surprising result. Economists agree that giving or donating behaviour as a function of income is not linear, but has a U-shaped pattern-people in the lowest and highest income groups and this relationship persists even when accounting for additional variables associated with income (McClelland and Brooks, 2004).[29] If so, in our case, we should observe negative coefficients on middle-family income students and a positive coefficient for respondents with a family income greater than 5,000 euros. Unfortunately, our findings satisfy only the first part of the story; indeed, the coefficient on the highest income group is negative rather than positive as predicted by the theory. Thus, this is probably more a problem of data rather than the consequence of an unexpected behaviour. The offered annual payments of EUR 0.5, 1 and 2 are very small, so everyone can afford them. Moreover, 43% of respondents voted the maximum option of EUR 2 (see above). Now, since there are few students in the survey who belong to a family earning more than EUR 5,000 per month (11.2%), the little variation in this category is the most plausible cause leading to a negative coefficient associated with this group.

---

[26] For example, Italy is the most represented country at the CERN with more than 700 scientists. Moreover at the end of 2014, the Italian physicist Fabiola Gianotti was elected General-Director of the CERN and her mandate began on 1 January 2016.
[27] The negative sign remains even when we use the availability of a personal income (unreported regression) rather than the family income confirming an inverse relationship between age and WTP, once income has been netted out.
[28] The non-significance of the variables "A1 Knowing what is a RI" and "A4 Interest in research" is due to the collinearity with "A2 Particle accelerator" and "A6 Importance in research" respectively. The Spearman's correlation coefficients are respectively 0.4 and 0.5 (see Table 5)

University degrees such as humanities and social sciences display negative coefficients with respect to scientific degrees. In the case of the highest category of WTP (column j= € 2), being enrolled in social science faculties reduces the probability of donating EUR 2 by 13 percentage points compared to students enrolled in scientific degrees such as mathematics or physics. Hence, although the variable education does not affect the participation equation (to pay or not to pay), it matters in determining the level of WTP.

As expected, importance of funding RI and having heard about LHC are associated with higher levels of WTP.

Finally, the taos ($\tau$) parameters refer to the thresholds used to differentiate the adjacent levels of the dependent variable. They are all statistically significant, justifying the use of three categories of the level of WTP over combining some categories. Moreover, the likelihood ratio tests in the models indicate that the variation in the independent variables explains a good proportion of the variability in the response variable.

The determinants of WTP when expressed in the form of a single lump-sum payment amounting to EUR 30 are investigated by estimating a standard multinomial (MNL) logit model. Estimation of a MNL model will enable us both to allow for and to test for differences in the factors associated with each of the three outcomes: '*yes*', '*no*' and '*do not know*'. Tests were also conducted to determine whether the assumption of Indipendence of Irrelevant Alternatives (IIA), underlying the MNL specification, holds.[30]

Multinomial logit estimates are reported in Table 8. As before, model 1 leaves out variables expressing attitude of students towards research; these factors were added in the model 2.

Table 8 reveals that the higher the income, the higher the probability of being willing to pay EUR 30 with respect to not be willing to pay (the base case), that is to say '*yes*' with respect to say '*no*' (Column 3). The effect of income is mitigated for students who answered '*do not know*' with respect to those who answered '*no*' (Column 4). Differently from income, the demographic characteristics (gender, age, education and household composition) are not significant determinants of WTP outcome.

The variable "A2 Particle accelerator" is positively associated with the WTP, i.e. respondents who identify a RI with a particle accelerator are more likely to answer '*yes*' than '*no*', as compared to respondents who did not recognise the LHC as a RI (Column 3). Moreover, this variable discriminates also '*do not know*' answers compared to '*no*' ones (Column 4).

In line with expectations, having some interest in research activities, judging funding research infrastructures important and having heard about LHC have jointly a positive and statistically significant impact on answering '*yes*' relative to '*no*'.

---

[29] A plausible explanation for this is that people in lower-income groups tend to give more to some organizations such as religious ones, while people in higher-income groups simply have more disposable income.

[30] According to the assumption of Indipendence of Irrelevant Alternatives (IIA), residuals must be independent each other, with a log-Weibull distribution. If two alternatives are more similar to one another than the third alternative, as might be supposed, for example, if individuals answering '*do not know*' and '*no*' behave similarly (Groothuis and Whitehead, 2002), we would expect the test of IIA to reveal such similarities. In order to test the IIA assumption, we conducted two Hausman tests. In the first one, the MNL results were compared with those from a binomial logit model between the '*no*' and '*yes*' samples. In the second one, the MNL results were compared with those from a binomial logit between the '*do not know*' and '*yes*' samples. The p-values associated with the resulting test statistics were 0.78 and 0.81 respectively, allowing us to fail to reject the IIA assumption and provide further credibility to the use of a MNL procedure.

**Table 7. Determinants of WTP. Ordered Logit Estimates: Models and Average Marginal Effects**

| | Model | | | | Marginal Effects on Predicted Probabilities. Model 2 | | | | | |
|---|---|---|---|---|---|---|---|---|---|---|
| | (1) | | (2) | | | | | | | |
| VARIABLES | coef | se | coef | se | j= € 0.5 | se | j= € 1 | se | j= € 2 | se |
| Family income | | | | | | | | | | |
|   1,000-3,000 | -0.43* | (0.25) | -0.44* | (0.25) | 0.03* | (0.02) | 0.06* | (0.03) | -0.09* | (0.05) |
|   3,000-5,000 | -0.70*** | (0.25) | -0.73*** | (0.26) | 0.06*** | (0.02) | 0.09*** | (0.03) | -0.16*** | (0.05) |
|   > 5,000 | -0.73** | (0.33) | -0.83** | (0.33) | 0.07** | (0.03) | 0.11*** | (0.04) | -0.18** | (0.07) |
| Female | -0.29* | (0.15) | -0.19 | (0.16) | 0.02 | (0.01) | 0.02 | (0.02) | -0.04 | (0.03) |
| Age | | | | | | | | | | |
|   26-30 | 0.04 | (0.24) | 0.02 | (0.24) | -0.00 | (0.02) | -0.00 | (0.03) | 0.01 | (0.05) |
|   31-35 | 0.66 | (0.55) | 0.63 | (0.59) | -0.05 | (0.04) | -0.08 | (0.07) | 0.13 | (0.11) |
|   > 35 | 0.43 | (0.77) | 0.21 | (0.78) | -0.02 | (0.06) | -0.03 | (0.10) | 0.05 | (0.16) |
| Education | | | | | | | | | | |
|   Humanities | -0.31* | (0.18) | -0.15 | (0.19) | 0.01 | (0.02) | 0.02 | (0.02) | -0.03 | (0.04) |
|   Social Science | -0.67*** | (0.21) | -0.59*** | (0.21) | 0.06** | (0.02) | 0.07*** | (0.02) | -0.13*** | (0.05) |
| Household composition | | | | | | | | | | |
|   3-5 | -0.06 | (0.23) | -0.11 | (0.24) | 0.01 | (0.02) | 0.01 | (0.03) | -0.02 | (0.05) |
|   > 5 | -0.36 | (0.38) | -0.28 | (0.40) | 0.03 | (0.04) | 0.03 | (0.05) | -0.06 | (0.09) |
| A1 Knowing what a RI is | | | 0.22 | (0.17) | -0.02 | (0.02) | -0.03 | (0.02) | 0.05 | (0.04) |
| A2 Particle accelerator | | | 0.20 | (0.17) | -0.02 | (0.02) | -0.03 | (0.02) | 0.04 | (0.04) |
| A4 Interest in research | | | 0.09 | (0.22) | -0.01 | (0.02) | -0.01 | (0.03) | 0.02 | (0.05) |
| A6 Importance of funding RI | | | 0.61*** | (0.12) | -0.06*** | (0.01) | -0.08*** | (0.01) | 0.13*** | (0.02) |
| B1 Having heard about LHC | | | 0.35** | (0.18) | -0.03** | (0.02) | -0.05* | (0.02) | 0.08** | (0.04) |
| B3 Having heard about H.Boson | | | 0.22 | (0.18) | -0.02 | (0.02) | -0.03 | (0.02) | 0.05 | (0.04) |
| $\tau_1$ | -3.18*** | (0.37) | 1.20* | (0.64) | | | | | | |
| $\tau_2$ | -1.40*** | (0.34) | 1.91*** | (0.65) | | | | | | |
| | | | | | | | | | | |
| Country –specific effects | yes | | yes | | | | | | | |
| Observations | 740 | | 740 | | 740 | | 740 | | 740 | |
| McFadden's R2 | 0.0229 | | 0.0618 | | | | | | | |
| Log Likelihood | -661.2 | | -634.3 | | | | | | | |
| Likelihood ratio test | 28.24 | | 73.38 | | | | | | | |

Table shows the determinants of the probability of falling in one of the WTP category. Robust standard errors in parentheses. ***,**,* denote significance at the 1%, 5% 10% level respectively

**Table 8. Determinants of WTP. Mutinomial logit estimates. Base outcome 'No'**

|  | Model 1 | | | | Model 2 | | | |
|---|---|---|---|---|---|---|---|---|
|  | (1) Yes | | (2) I do not know | | (3) Yes | | (4) I don not know | |
| VARIABLES | Coef | se | coef | se | coef | se | coef | se |
| Family income | | | | | | | | |
|    1,000 - 3,000 | 0.56* | (0.30) | 0.34 | (0.22) | 0.49* | (0.30) | 0.18 | (0.22) |
|    3,000 – 5,000 | 0.63** | (0.32) | 0.45** | (0.23) | 0.59* | (0.34) | 0.34 | (0.24) |
|    > 5,000 | 1.20*** | (0.39) | 1.02*** | (0.30) | 1.01** | (0.41) | 1.01*** | (0.30) |
| Female | -0.11 | (0.21) | 0.04 | (0.16) | 0.15 | (0.22) | 0.05 | (0.16) |
| Age | | | | | | | | |
|    26-30 | 0.904*** | (0.33) | 0.34 | (0.26) | 0.84** | (0.34) | 0.29 | (0.26) |
|    31-35 | -0.56 | (0.67) | 0.02 | (0.39) | -0.25 | (0.65) | -0.08 | (0.40) |
|    > 35 | -0.16 | (0.59) | -1.24 | (0.68) | -0.06 | (0.60) | -1.26* | (0.68) |
| Education | | | | | | | | |
|    Humanities | -0.13 | (0.24) | 0.03 | (0.17) | 0.17 | (0.25) | 0.11 | (0.18) |
|    Social Science | -0.19 | (0.25) | -0.27 | (0.20) | -0.03 | (0.26) | -0.24 | (0.20) |
| Household composition | | | | | | | | |
|    3-5 | -0.43 | (0.33) | -0.31 | (0.24) | -0.55 | (0.34) | -0.34 | (0.25) |
|    > 5 | -0.51 | (0.47) | -0.26 | (0.33) | -0.30 | (0.49) | -0.25 | (0.33) |
| A1 Knowing what a RI is | | | | | -0.29 | (0.23) | -0.09 | (0.17) |
| A2 Particle accelerator | | | | | 0.66*** | (0.23) | 0.40*** | (0.17) |
| A4 Interest in research | | | | | 0.55* | (0.33) | 0.47** | (0.21) |
| A6 Importance of funding RI | | | | | 0.49*** | (0.18) | 0.05 | (0.10) |
| B1 Having heard about LHC | | | | | 0.43* | (0.24) | -0.18 | (0.18) |
| B3 Having heard about Higgs Boson | | | | | 0.25 | (0.26) | 0.10 | (0.25) |
| Constant | -0.02 | (0.50) | 0.17 | (0.34) | -3.28*** | (1.04) | -0.41 | (0.55) |
| | | | | | | | | |
| Country-specific effects | Yes | | yes | | yes | | yes | |
| Observations | 1,010 | | 1,010 | | 1,009 | | 1,009 | |
| % Correct predictions | 50.2 | | | | 52.1 | | | |
| McFadden's R2 | 0.089 | | | | 0.114 | | | |
| Log Likelihood | -914.1 | | | | -889.0 | | | |
| Likelihood ratio test | 179.3 | | | | 228.2 | | | |

Robust standard errors in parenthesis. ***,**,* denote significance at the 1%, 5% 10% level respectively.

Summing up, our empirical analysis does not show significant differences between the explanatory variables associated with the WTP, when the WTP is investigated either by asking for a one-time payment or by asking for a recurring annual contribution. Income and variables expressing interest and attitude towards research are positively associated with a 'yes' response in both cases; in contrast and conditional to our sample, the type of university degree does not discriminate between paying and not paying. The type of education matters, however, in the WTP level equation: students enrolled in scientific degrees such as mathematics, physics, engineering and medicine are willing to pay more for science. In contrast to the expectations, income shows a negative sign in the level equation, a result probably due to the fact that most respondents do not earn income on their own.

### 5.2 *The truncated mean WTP*

Results of the logit model (11) are presented in Table 9. As expected, the estimated coefficient on BID was found negative and statistically significant. Family income significantly impacts on the probability of WTP '*yes*' and the sign was positive. The effect of other variables was discussed in the previous sections. The Count $R^2$ reveals that 75.5% of observations were correctly allocated to predict WTP either '*yes*' or '*no*', indicating a good fit to the data.

**Table 9. Estimation of WTP. Results of logit models.**

| VARIABLES | (1) Coef | se | (2) dy/dx | se |
|---|---|---|---|---|
| BID | -0.481* | (0.282) | -0.081* | (0.047) |
| Family Income | 0.174* | (0.102) | 0.029* | (0.017) |
| Female | 0.020 | (0.204) | 0.003 | (0.034) |
| Age | | | | |
|    26-30 | 1.258*** | (0.321) | 0.227*** | (0.058) |
|    31-35 | -1.079** | (0.567) | -0.159** | (0.069) |
|    >35 | -0.738 | (0.590) | -0.114 | (0.082) |
| Household composition | 0.031 | (0.224) | 0.005 | (0.038) |
| Education | -0.065 | (0.112) | -0.010 | (0.019) |
| A2 Particle Accelerator | 0.667*** | (0.202) | 0.114*** | (0.034) |
| A6 Importance of funding RI | | | | |
|    Important enough | 0.843 | (0.736) | 0.133 | (0.106) |
|    Important | 1.306* | (0.693) | 0.220* | (0.127) |
|    Fundamental | 1.657*** | (0.691) | 0.315*** | (0.124) |
| B1 Having heard about LHC | 0.366* | (0.206) | 0.062* | (0.033) |
| B3 Having heard about Higgs Boson | 0.436* | (0.235) | 0.073* | (0.039) |
| Constant | -0.892 | (0.929) | | |
| | | | | |
| Country-specific effects | Yes | | | |
| Observations | 740 | | | |
| Count R2 | 75.5 | | | |
| McFadden's R2 | 0.208 | | | |
| Log Likelihood | -364.8 | | | |
| Likelihood ratio test | 191.29 | | | |

Robust standard errors in parenthesis. ***,**,* denote significance at the 1%, 5% 10% level respectively. (*) dy/dx is for discrete change of dummy variable from 0 to 1.

Once the parameters of the logit model are estimated, the truncated mean WTP is given by

$$E(WTP) = \int_0^{30} [1 + \exp(-(3.66 - 0.481A))]^{-1} \delta A = €\ 7.7 \qquad (12)$$

The estimated mean WTP for basic research at the LHC is EUR 7.7 per person.

## 6. Concluding remarks

Large scale research infrastructures are costly, and governments are often asking to what extent they should fund them, particularly in the domain of basic science. This paper contributes to the literature on the welfare effects of science in several ways. First, we have proposed that beyond use-benefits of science, such as those identified by Florio and Sirtori (2015), there are non-use benefits. There is indirect evidence of such benefits through the wide cultural effects of discovery announcements, such as about the Higgs boson (or more recently about the gravitational waves also widely reported in the media). Recurring to the concept of existence-value, in analogy with environmental and cultural economics, allows us to propose a direct empirical analysis of the non-use value of science for the general public.

Our second contribution is to study the determinants of the WTP for basic science, pointing to the LHC as an example of a highly visible research infrastructure, that generates knowledge not yet associated to any predictable application. We show how this analysis of the determinants can be implemented, and we think our approach is easily replicable for other research infrastructures and scientific programmes.

Finally, we estimate, for the first time, the WTP for basic research of a relatively large sample of respondents (comparable or greater then in most earlier CV literature), by using methods largely applied in welfare economics. Again, we suggest that replication with larger samples of the general population along the lines we have presented is feasible (even if costly because we advocate personal in depth interviews).

We conclude with some caveats and indications for future research. The estimated WTP for the LHC case is certainly on the lower side of the actual distribution for several reasons mentioned in the paper, and hence cannot be taken at face value. However, by applying our estimation to the adult population of CERN member states, which amounted to 481 million people in 2013[31], it would suggest a non-use social benefit of EUR 3.7 billion (USD 4.1 billion). By adding to this population an additional 21%[32] to take into account the population of CERN non member states that directly or indirectly support the LHC (e.g. notably the USA), the target population increases to 610.5 million, and the non-use benefit rises to EUR 4.7 billion (USD 5.1 billion). These figures are in the same range of the estimated non-use value provided by Florio *et al.* (2016a) who use alternative approaches to CV and, whitin their framework, contributes substantially to achieve a positive social net present value of the LHC to 2025. This suggests that ignoring the non-use value in cost benefit analysis of research infrastructures would lead to a severe underestimation of their overall socio-economic impact.

---

[31] Our elaborations on Florio *et al.* (2016a) data. The target population amounts to 481 million people aged between 18 and 80 in the following member states: Austria, Belgium, Bulgaria, Czech Republic, Denmark, Finland, France, Germany, Greece, Hungary, Italy, Netherlands, Norway, Poland, Portugal, Slovak Republic, Spain, Sweden, Switzerland, United Kingdom, Israel

[32] The percentage of 21% reflects the share of onsite visitors to CERN from non-member states. For further details see Florio *et al.* (2016a).

# ANNEX I - QUESTIONNAIRE

# CONTINGENT VALUATION
# OF A RESEARCH INFRASTRUCTURE

| *University:* | *Faculty:* |
|---|---|

| SECTION A | |
|---|---|
| **A.1.** Do you know what a research infrastructure is? | □ YES<br>□ NO |
| **A.2.** In your opinion, which of the following is a research infrastructure?<br><br>*For this question it is possible to choose multiple answers. | □ TELESCOPE<br>□ INSTRUMENT OF DATA COLLECTION AND ARCHIVE<br>□ DATA ELABORATION SOFTWARE<br>□ PARTICLE ACCELERATOR<br>□ LIBRARY<br>□ COMPUTER<br>□ ASTRONOMICAL OBSERVATORY<br>□ PLANETARIUM |
| **A.3.** Can you give an example of a research infrastructure that you know/that you visited or that you heard of? | |
| **A.4.** Are you interested in scientific discoveries and in research activities in general? | □ YES<br>□ NO |
| **A.5.** If yes, please indicate your source of information. | □ TV<br>□ Radio<br>□ Specialised magazines<br>□ Online news<br>□ Other (please specify) |
| **A.6.** On a scale from 1 (not relevant) to 5 (essential), how do you rate the importance of funding research infrastructures? | □ Useless<br>□ Insignificant<br>□ Important enough<br>□ Important<br>□ Fundamental |
| **A.7.** Can you briefly explain the motivations for your answer on the previous question? | |

## *LARGE HADRON COLLIDER (LHC)*

*What is the LHC?*

The Large Hadron Collider (LHC) is the biggest and most powerful particle accelerator ever built. It can accelerate adrons (heavy protons and iones) up to 99,9999991% of speed of light and make them collide afterwards, currently reaching an amount of energy, in the mass centre, of 8 teraelettronvolt (it is expected that, in 2015, this energy will reach near 14 teraelettronvolt, which is the full capacity of the infrastructure).
It's located at the CERN of Geneva and is built inside an underground 27 km-long tunnel located on the border between France and Switzerland, in a region included between the Geneva airport and the Giura mountains, originally excavated to build the Large Electron-Positron Collider (LEP).
The tunnel is located at an average depth of 100 meters.
It is formed by about 2,000 superconductive magnets, maintained at a temperature of -271 °C. The low temperature serves to create in the magnets the phenomenon called 'superconductivity': this way much less energy is consumed and it is possible to accelerate particles at high energies. The machine accelerates two particle beams circulating in opposite directions, each of them contained in a vacuum tube. Those then collide in four points along the orbit, in correspondence to caverns where the tunnel widens in large experimental halls. In these stations there are four principal experiments of particle physics: ATLAS (A Toroidal LHC Apparatus), CMS (Compact Muon Solenoid), LHCb (LHC-beauty) and ALICE (A Large Ion Collider Experiment). Those enormous facilities consist in a large number of detectors that use different technologies and operate around the pint where the beams collide. During collisions, thanks to the transformation of a part of the high energy in mass, a very large amount of particles is produced whose properties are measured by the detectors. The two smallest detectors are TOTEM and LHCf.

The entry into operation of LHC, originally scheduled for the ned of 2007, took place on September 10, 2008. Italy participates in the LHC project within the context of its contribution to CERN as member state (the division of CERN budget between member Countries is based on the GDP).

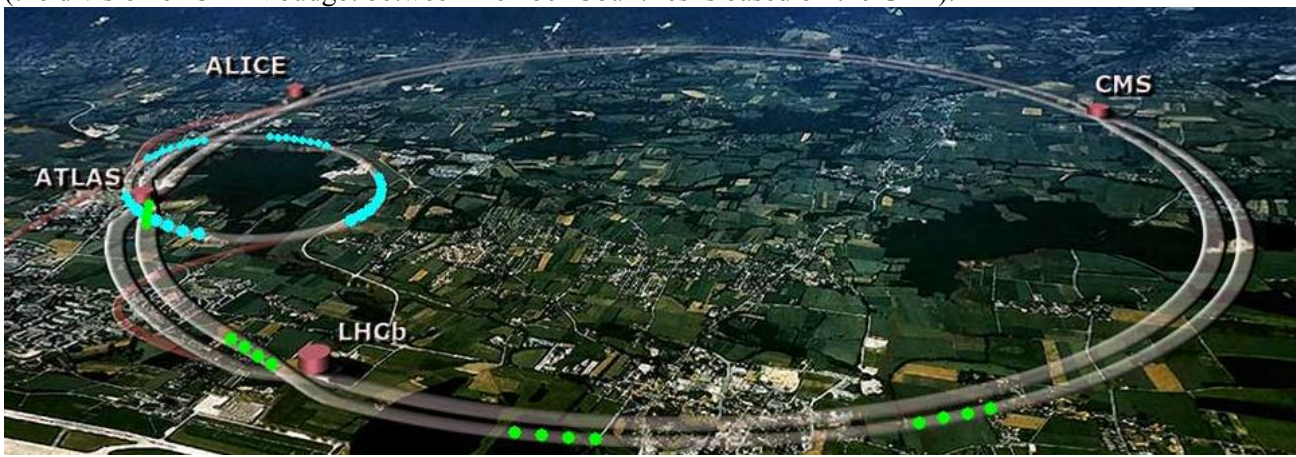

*What is the purpose of LHC and its single experiments?*

LHC is used for experimental research in the field of particle physics. In particular, LHC serves to discover what the vast majority of matter and energy contained in the universe is made of. Today we only know that there exists a lot of dark matter and dark energy, but we do not know what they are made of. In July 2012 LHC reached its first big achievement: it "saw" the famous Higgs' Bosons, the particle whose field allows all the particles to have a mass. This particle helps explain why mass exists. Each experiment plays a specific research activity:

- **ATLAS e CMS**: revealed the Higgs Boson and deal with the research of supersimmetry.
- **ALICE**: works on plasma of Quark and the gluons, a state of matter existed in the first moments after the Big Bang.
- **LHC-b**: studies how asimmetry between matter and antimatter was created.
- **LHC-f**: it is the smallest experiment and verifies theories on cosmic rays.

- **TOTEM**: Measure the probability and the impact modalities between protons in the LHC.

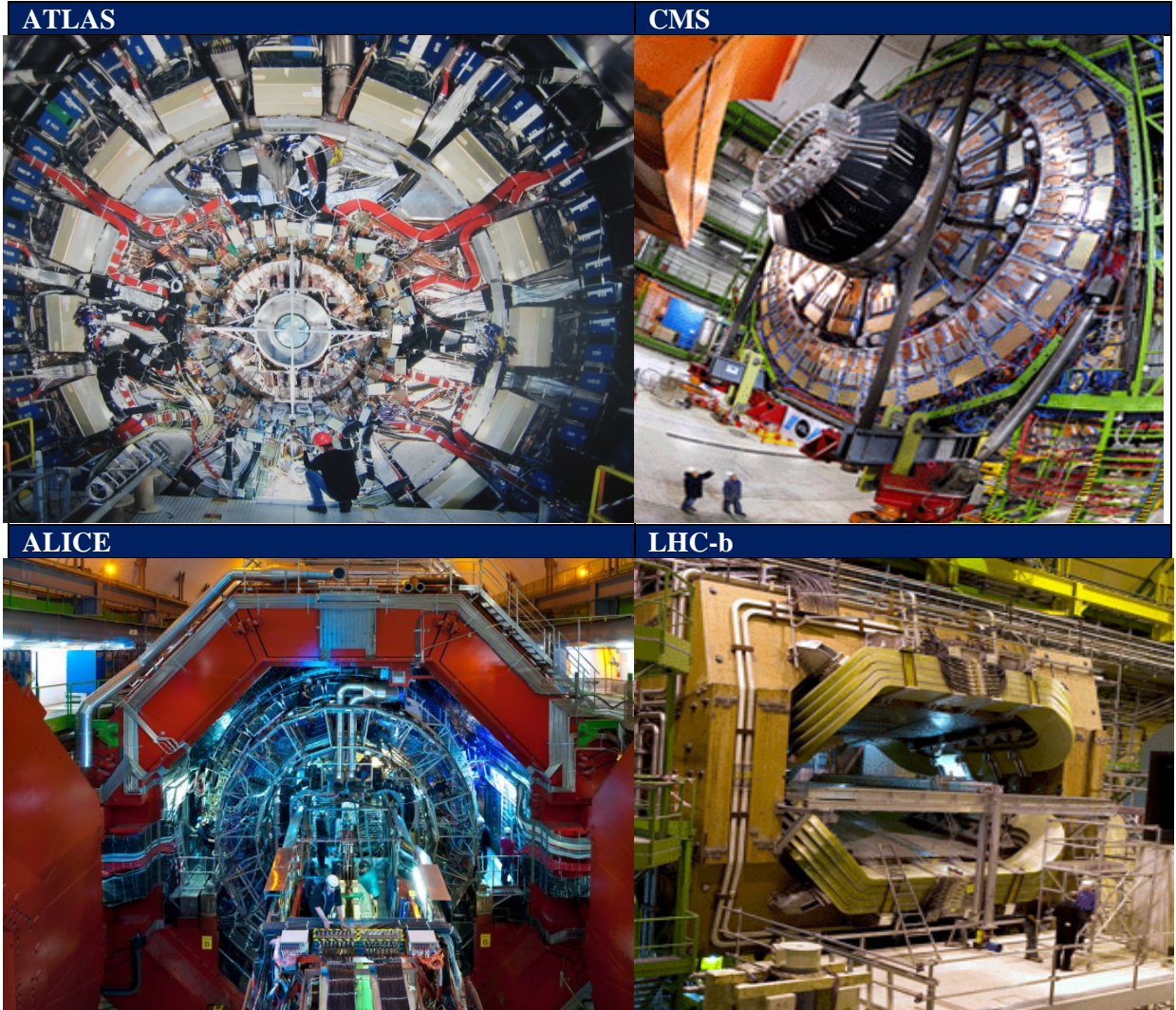

In future, LHC may also discover the existence of supersimmetrycal particles and push us to think that universe is not made of only four dimensions as we perceive (right-left, up-down, back-forth, plus the time dimension) but by many other dimensions invisible to us, rolled up on themselves. LHC may help us understand:

- ✓ Why the matter we are made of is so stable during the time;
- ✓ Why universe is expanding faster than expected;
- ✓ Why universe seems to be made by a 95% of something we do not see and we do not know, but acts on galaxies and therefore exists.

**SOURCE:** *Wikipedia (28/05/2014)*

| SEZIONE B | |
|---|---|
| **B.1.** Did you hear about the LHC before this questionnaire? | □ YES<br>□ NO |
| **B.2.** If yes, please indicate your source of information. | □ School/university<br>□ TV<br>□ Magazines<br>□ Internet<br>□ Friends<br>□ Other (please specify) |
| **B.3.** Did you ever hear of "Higgs Boson"? | □ YES<br>□ NO |
| **B.4.** If yes, please indicate your source of information. | □ School/university<br>□ TV<br>□ Magazines<br>□ Internet<br>□ Friends<br>□ Other (please specify) |
| **B.5.** Have you ever been to the CERN? | □ YES<br>□ NO |
| **B.6.** If yes, please indicate how many times? | □ once<br>□ twice<br>□ more than twice |
| **B.7.** In your opinion, what is the purpose of the LHC?<br><br>*For this question it is possible to choose multiple answers. | □ it is an useless machine whose construction could have been avoided.<br>□ it is a machine that affects only physicians' and scientists' world.<br>□ it is a dangerous machine for the risk of nuclear accidents.<br>□ it is a useful machine for energy production.<br>□ it is a useful machine for experiment with protons' acceleration which can be used for different purposes. |
| **B.8.** Would you be willing to provide an economic contribution to fund the research activity of the LHC? | □ YES<br>□ NO<br>□ I DON'T KNOW |
| **B.9.** Could you please explain why you would be **(or WOULDN'T be)** willing to fund the research activity of the LHC? | |
| **B.10.** By 2015, would you be willing to offer an economic contribution equal to 30 Euro (lump sum), turning down other personal expenses? | □ YES<br>□ NO<br>□ I DON'T KNOW |
| **B.11.** Could you please explain why would you be **(or WOULDN'T be)** willing to pay a sum equal to 30 euro lump sum? | |
| **B.12** If someone asks you to give an economic contribution to the LHC by means of an annual tax over a period of 30 years, would you be willing to pay an annual amount equal to: | □ 2 EURO<br>□ 1 EURO<br>□ 0.50 EURO<br>□ 0 EURO |
| **B.13.** Could you please explain why you would be willing to pay this contribution? | |

| SEZIONE C | |
|---|---|
| **C.1.** Age: | □ 19-25<br>□ 26-30<br>□ 31-35<br>□ more than 35 years old |
| **C.2.** Sex: | □ M<br>□ F |
| **C.3.** City of residence: | |
| **C.4.** What is your educational background (pre-university): | □ scientific<br>□ classical<br>□ technical<br>□ Other (please specify) |
| **C.5** What was your average score during your pre-university studies? | Pass (50-59)<br>Merit (60-69)<br>Distinction (>70) |
| **C.6.** In which of the following brackets does your family monthly net income fall? | □ up to 1,000 Euro<br>□ from 1,000 to 3,000 Euro<br>□ from 3,000 to 5,000 Euro<br>□ more than 5,000 |
| **C.7.** Do you have your own personal income? | □ YES<br>□ NO |
| **C.8.** Household composition (including parents, brothers/sisters): | □ 1-2<br>□ from 3 to 5<br>□ more than 5 |